\documentclass[aps,prl,reprint,amsmath,amssymb,showpacs,floatfix]{revtex4-1}

\usepackage{color,graphicx}

\begin{document}

\title{Excitons and biexcitons in symmetric electron-hole bilayers}

\author{Ryo Maezono}
\affiliation{
  School of Information Science, JAIST,
  Asahidai 1-1, Nomi, Ishikawa, 923-1292, Japan}

\author{Pablo \surname{L\'opez R{\'\i}os}}
\affiliation{
  Theory of Condensed Matter Group,
  Cavendish Laboratory,
  J J Thomson Avenue, Cambridge CB3 0HE, UK}

\author{Tetsuo Ogawa}
\affiliation{
  Department of Physics,
  Osaka University,
  Toyonaka, Osaka 560-0043, Japan}

\author{Richard J.\ Needs}
\affiliation{
  Theory of Condensed Matter Group,
  Cavendish Laboratory,
  J J Thomson Avenue, Cambridge CB3 0HE, UK}

\date{\today}

\begin{abstract}
  Symmetric electron-hole bilayer systems have been studied at zero
  temperature using the diffusion quantum Monte Carlo method.
  A flexible trial wave function is used that can describe fluid,
  excitonic and biexcitonic phases.
  We calculate condensate fractions and pair correlation functions
  for a large number of densities $r_s$ and layer separations $d$.
  At small $d$ we find a one-component fluid phase, an excitonic
  fluid phase, and a biexcitonic fluid phase, and the transitions
  among them appear to be continuous.
  At $d=0$, excitons appear to survive down to about $r_s=0.5$~a.u.,
  and biexcitons form at $r_s>2.5$~a.u.
\end{abstract}


\pacs{02.70.Ss, 71.35.-y, 71.10.-w}

\maketitle


Electron-hole bilayer systems in which electrons and holes are
generated via doping and confined to separate layers by the
application of an electric field have been developed in, for example,
GaAs/AlGaAs heterostructures.\cite{Gupta_2011}
Other systems have been investigated, such as electron-hole bilayers
with very small electron-hole separations at oxide
interfaces\cite{Pentcheva_2010,Huijben_2012} and bilayer graphene
systems with approximately equal electron and hole
masses.\cite{Perali_2013}
These systems are expected to exhibit rich phase diagrams with Fermi
fluid, excitonic superfluid, biexcitonic, and charge density wave
phases.\cite{DePalo_2002,Senatore_2003,Schleede_2012}
We have studied the simplest possible such model system, with equal
electron and hole populations and equal masses, and parallel
infinitely-thin two-dimensional layers of variable separation and
carrier density.
It is important to establish the behavior of this simple system before
more complicated cases such as those of unequal electron and hole
masses\cite{Subasi_2010} and/or unequal electron and hole
densities\cite{Parish_2011} can be tackled with confidence.
Further work will be required to study more realistic systems with
anisotropic masses, finite well widths and depths, interface
roughness, etc.
Theoretical studies of correlation effects in electron-hole bilayers
have used methods such as dielectric
formulations,\cite{Moudgil_2002,Singwi_1968,Golden_2012}
Bardeen-Cooper-Schrieffer theory,\cite{Subasi_2010} and diffusion
\cite{DePalo_2002,Senatore_2003} and path integral
\cite{Schleede_2012} quantum Monte Carlo methods.

We consider paramagnetic, symmetric, electron-hole bilayers consisting
of $N$ up- and down-spin electrons and holes of equal masses,
$m_e=m_h$, where the distance between the two parallel layers is $d$.
Hartree atomic units are used throughout
($\hbar=|e|=m_e=4\pi\epsilon_0=1$).
The Hamiltonian of the infinite system is
\begin{eqnarray}
\label{eq:hamiltonian}
\nonumber
\hat H &=& -\frac 1 2 \left( \sum_{i} \nabla_{{\bf e}_i}^2 +
                             \sum_{i} \nabla_{{\bf h}_i}^2 \right)
           + \sum_{i<j} \frac{1}{\left|{\bf e}_i-{\bf e}_j\right|} \\
       & & + \sum_{i<j} \frac{1}{\left|{\bf h}_i-{\bf h}_j\right|}
           - \sum_{i,j} \frac{1}{
               \sqrt{d^2+\left|{\bf e}_i-{\bf h}_j\right|^2}}
           \;,
\end{eqnarray}
where ${\bf e}_i$ and ${\bf h}_j$ are the in-plane position vectors of
the $i$th electron and the $j$th hole.
We use finite simulation cells subject to periodic boundary conditions,
and the Coulomb sums are evaluated using two-dimensional Ewald
sums.\cite{Parry_1975}

Our results have been obtained with $N=29$ particles of each type,
giving a total of 116 particles, although we have also simulated the
system with $N=57$, corresponding to 228 particles, to investigate
finite size effects, which we find to be small.
The parameters that define the system are $d$ and the in-layer density
parameter $r_s=a/\sqrt{2\pi N}$, where $a$ is the side of the
square simulation cell.
The $d$ parameter controls the interaction between layers, while $r_s$
controls the interaction within the layers.
In this paper we focus on the density range $r_s<10$~a.u., and we
have not considered the very low density regime within which Wigner
crystallization is favorable.\cite{DePalo_2002,Senatore_2003}
At large $d$ the electron and hole layers become decoupled and the
results for each layer tend towards those for the two-dimensional
electron gas,\cite{Tanatar_1989,Attaccalite_2002,Drummond_2d_HEG_2009}
and when inter-layer and intra-layer interactions become comparable,
i.e., at $d \lesssim r_s$, a paired phase is expected.
It has been shown that biexciton formation is energetically favorable
at low densities when $d<0.38$ a.u.\cite{Schindler_2008,Lee_2009}
Biexciton formation is expected to be suppressed at high densities,
and this has been estimated to occur for $r_s \lesssim
10$~a.u.\cite{Zhu_1996_Physica_Scripta}

We have used the variational and diffusion quantum Monte Carlo (VMC
and DMC) methods as implemented in the \textsc{casino}
code.\cite{casino_reference}
Expectation values are obtained with VMC by importance sampled Monte
Carlo integration using an importance distribution $|\Psi_{\rm T}|^2$,
where $\Psi_{\rm T}$ is a suitable trial wave function.
$\Psi_{\rm T}$ contains a number of optimizable parameters whose
values are fixed by optimization at each $d$ and $r_s$.
DMC is a projector method in which expectation values are computed by
approximate solution of the imaginary-time-dependent Schr\"odinger
equation.\cite{ceperley_1980,foulkes_rmp_2001}
We use the standard ``fixed-node'' approximation to maintain the
fermionic symmetry of the system.\cite{anderson_1975,anderson_1976}
DMC expectation values are typically more accurate than those from
VMC, and in particular the accuracy of DMC energies only depends on
the quality of the nodal surface of $\Psi_{\rm T}$.
We use the standard mixed estimator to evaluate the DMC expectation
values reported in this work.\cite{foulkes_rmp_2001}

In the DMC study of electron-hole bilayers by De Palo
\textit{et al.},\cite{DePalo_2002,Senatore_2003} each phase of the
system was described by a different wave function, and the relative
stability of the phases was determined using the total energy.
In our study we use a single flexible wave function form which is
capable of describing the Fermi liquid, excitonic superfluid, and
biexcitonic phases, and the character of the system at each $r_s$ and
$d$ is investigated by computing the expectation values of the
electron-hole condensate fraction and the pair-correlation functions
(PCFs).

We have used a Slater-Jastrow (SJ) trial wave function,
\begin{equation}
\label{eq:wave_function}
\Psi_{\rm T} =
 \exp\left[J({\bf R})\right]
 \det\left[ \phi({\bf e}_i^\uparrow-{\bf h}_j^\downarrow) \right]
 \det\left[ \phi({\bf e}_i^\downarrow-{\bf h}_j^\uparrow) \right] \;,
\end{equation}
where $\exp\left[J({\bf R})\right]$ is a Jastrow correlation factor
that depends on all of the particle positions ${\bf R}$, and the
pairing orbitals are
\begin{equation}
\label{eq:mixed_pairing}
\phi({\bf r}) =
  \sum_{l=1}^{n_{\rm p}} p_l \cos({\bf k}_l \cdot {\bf r}) +
  f(r;L) \sum_{m=0}^{n_{\rm c}} c_m r^m \;,
\end{equation}
where $n_{\rm p}$ is the plane-wave expansion order,
${\bf k}_l$ is the $l$th shortest reciprocal-space vector,
$n_{\rm c}$ is the polynomial expansion order, $f(r;L)$ is a
cut-off function given by $f(r;L)=\left(1-r/L\right)^3
\Theta\left(r-L\right)$, $\Theta$ is the Heaviside step function,
and $\{p_l\}$, $\{c_m\}$, and $L$ are optimizable parameters.
We constrain $p_l=p_{l^\prime}$ whenever $\left|{\bf k}_l\right|$ and
$\left|{\bf k}_{l^\prime}\right|$ are in the same star.
This form describes a pure fluid phase when $n_{\rm p}=N$, $p_l\neq 0$
for all $l$, and $c_m=0$ for all $m$, and an excitonic phase when
$p_l=0$ for all $l$.
This wave function cannot describe biexcitons since it only binds
antiparallel-spin electron-hole pairs, and biexciton correlations are
introduced by the Jastrow factor.

We have used a Drummond-Towler-Needs Jastrow
factor\cite{drummond_jastrow_2004} consisting of a two-body polynomial
$u$ term, to which the electron-electron, hole-hole, and electron-hole
Kato cusp conditions are applied.\cite{kato_1957}
The electron-hole cusp condition is only applicable when $d=0$, which
makes it difficult to obtain results of the same degree of accuracy
for $d=0$ and $d>0$.
To solve this problem we have introduced a ``quasi-cusp'' Jastrow
factor term, $Q$, which smoothly introduces the electron-hole cusp
condition as $d\rightarrow 0$, but we do not use it when $d=0$ since
the $u$ term enforces the exact cusp.\cite{supplemental}
The $Q$ term contains a single optimizable cut-off length.

We have used expansion orders of $n_{\rm p}=81$ (14 stars of
$k$-vectors) and $n_{\rm c}=8$.
Our wave function contains a total of 47 optimizable parameters
at $d=0$, and 48 at $d>0$.
We have optimized these parameters within VMC using linear
least-squares energy
minimization.\cite{toulouse_emin_2007,umrigar_emin_2007}

\begin{figure}[htb!]
  \begin{center}
    \includegraphics[width=0.36\textwidth]{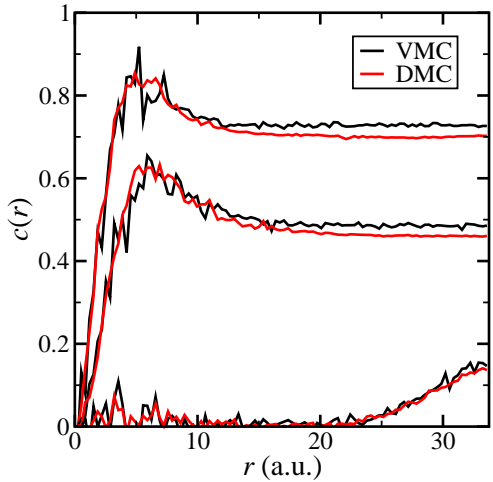}
  \end{center}
  \caption{
    (Color online)
    VMC and DMC expectation values of $c(r)$ for $r_s=5$~a.u.\ and
    (from top to bottom) $d=0.3$, $1$, and $4$~a.u.
    The systems with $d=0.3$ and $1$~a.u.\ are in the excitonic phase,
    and that with $d=4$~a.u.\ is in the fluid phase.}
  \label{fig:tbdm_rs5}
\end{figure}

The translational-rotational average of the two-body
density matrix for electron-hole pairs is
\begin{equation}
\label{eq:2bdm_TR}
\rho_{eh}^{(2)}(r) =
  \frac {N^2 \int \left| \Psi({\bf R}) \right|^2
         \frac {\Psi({\bf e}_1+{\bf r}^\prime,
                     {\bf h}_1+{\bf r}^{\prime})}
               {\Psi({\bf e}_1,
                     {\bf h}_1)}
         \delta(|{\bf r}^\prime|-r) d{\bf R} d{\bf r}^\prime }
        {\Omega^2 2\pi r
         \int \left| \Psi({\bf R}) \right|^2 d{\bf R} } \;,
\end{equation}
where $\Omega$ is the area of the simulation cell.
The condensate fraction $c$ is defined as the large-$r$ limit of
$\rho_{eh}^{(2)}(r)$ normalized so that $c=1$ when all electrons and
holes are bound into excitons.\cite{yang_1962}
We have evaluated $c$ using the improved estimator of
Ref.~\onlinecite{Astrakharchik_2005}, which we call $c(r)$, see
Fig.~\ref{fig:tbdm_rs5}.
The condensate fraction is zero for pure one-component and
biexcitonic fluid phases.

We also compute the translational-rotational average of the PCF,
\begin{equation}
\label{eq:pcf}
g_{\alpha\beta}(r) =
  \frac{\Omega
        \int |\Psi({\bf R})|^2 \delta\left({\bf r}_\alpha-
             {\bf r}_\beta-{\bf r}^\prime\right)
             \delta\left(\left|{\bf r}^\prime\right|-r\right)
             d{\bf R} d{\bf r}^\prime}
       {2\pi r
        \int |\Psi({\bf R})|^2 d{\bf R}} \;,
\end{equation}
where $\alpha$ and $\beta$ are indices that distinguish the four
particle types in the system (up- and down-spin electrons and holes).
The PCFs allow us to detect biexciton formation, distinguishing the
biexcitonic phase from the one-component fluid, for both of which
$c=0$.

We used a target walker population of 1280 configurations and a time
step of $0.01$~a.u.\ for the DMC calculations.
We verified that the energy, condensate fraction, and PCF do not
change significantly when the time step was reduced from this
value.
The accuracy of a trial wave function can be measured by the
differences between expectation values calculated with the VMC and DMC
methods.
We find that these differences are small.
To investigate the convergence of our results with respect to the
quality of the wave function, we have also performed calculations
using a more sophisticated Slater-Jastrow-backflow (SJB) wave
functions for selected cases.
These wave functions incorporate a backflow transformation in which
the particle coordinates are replaced by ``quasiparticle''
coordinates,\cite{Kwon_1993_backflow,Lopez-Rios_2006} which adds 27
optimizable parameters to the wave function.
The introduction of backflow results in significant changes in the
computed expectation values at small values of $d$ but, as $d$
increases, the difference declines.
This indicates that the description of in-layer correlations afforded
by the SJ wave function is very good, while the description of
correlations between the motion in the electron and hole layers is not
as good.

We have computed $c(r)$ within VMC and DMC, and have evaluated the
condensate fractions as the average of $c(r)$ over the region of the
plateau at large $r$.\cite{supplemental}
Three examples of $c(r)$ functions are shown in
Fig.~\ref{fig:tbdm_rs5}.
The VMC and DMC values of the condensate fraction differ by less than
$3$\% in each case.
In the fluid phase $c(r)$ is close to zero for small values of $r$,
but it rises in value as $r$ reaches the edge of the simulation cell.
We interpret this as an effect due to the finite size of the simulation
cell, and take $c$ to be zero when this feature is present.

\begin{figure}[htb!]
  \centering
  \includegraphics[width=0.36\textwidth]{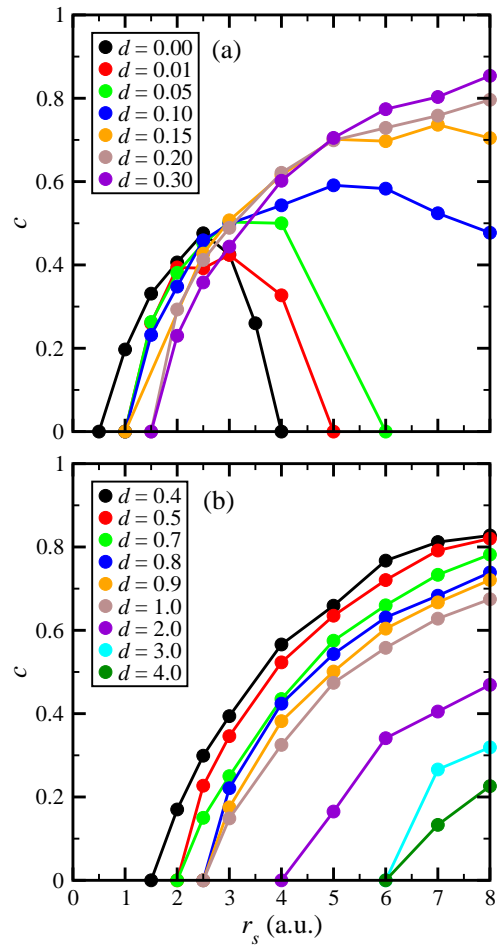}
  \caption{
    (Color online)
    DMC condensate fraction as a function of $r_s$ at
    (a) $d< 0.4$ a.u.\ and
    (b) $d\geq 0.4$ a.u.}
  \label{fig:cfrac_dmc}
\end{figure}

\begin{figure}[htb!]
  \centering
  \includegraphics[width=0.45\textwidth]{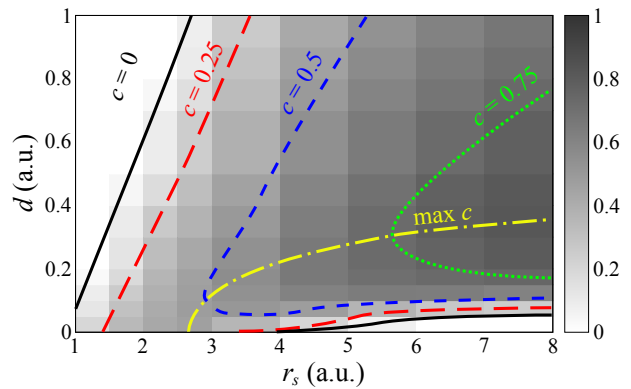}
  \caption{
    (Color online)
    The estimated DMC condensate fraction as a function of $r_s$ and
    $d$.
    Phase boundaries are represented by solid lines, and contours of
    the condensate fraction are shown as long-dashed, short-dashed,
    and dotted lines.
    The dot-dashed line indicates the location of the maximum $c$ for
    each $r_s$, below which some degree of biexciton formation takes
    place.}
  \label{fig:cfrac_dmc_map}
\end{figure}

Our results for the condensate fraction agree with those of
Ref.\ \onlinecite{DePalo_2002} for large $d$, but we tend
to obtain larger condensate fractions for small $d$.
Our main results for the condensate fractions are shown in Figs.\
\ref{fig:cfrac_dmc} and \ref{fig:cfrac_dmc_map}.
%
%
%
Fig.\ \ref{fig:cfrac_dmc}(a) shows that for small values of $d$ and
$r_s\geq 3$~a.u.\ the condensate fraction curves fall to zero with
increasing $r_s$, which can be attributed to the formation of
biexcitons.
%
%
Biexciton formation is favorable only at small $d$, because at large
$d$ the in-plane repulsion between like charges dominates the weak e-h
attraction.
Fig.~\ref{fig:cfrac_dmc_map} shows the condensate fraction as a
function of $r_s$ and $d$, including smoothed phase boundaries and
other contour lines, and a line that locates the maximum $c$ for each
$r_s$.
Since biexciton formation is the only likely mechanism by which $c$
can be reduced as $d$ decreases, this line delimits the region where
biexciton formation takes place.
The maximum condensate fraction for large values of $r_s$ occurs at
$d=0.4$ a.u., and $c$ increases with $r_s$ reaching, for instance,
$c=0.95$ at $r_s = 15$~a.u.

Studies of the bilayer system with two anti-parallel-spin electrons
and holes have shown that biexciton formation is energetically
favorable for $d<0.38$ a.u.\cite{Schindler_2008,Lee_2009}
We have found that biexciton formation is favorable in extended
systems at large $r_s$ for values of $d$ similar to those for the
isolated biexciton,\cite{Schindler_2008,Lee_2009} but that
biexcitons do not form below $r_s \simeq 2.5$~a.u., see
Fig.\ \ref{fig:cfrac_dmc}.

Condensate fractions calculated using a bosonic dipole model have been
reported in the
literature.\cite{Astrakharchik_PRL_2007,Astrakharchik_PRA_2007,Filinov_2012}
The behaviour of this bosonic system with a repulsive interaction
differs qualitatively from that of the electron-hole bilayer model at
small $d$, since the repulsive interaction is not capable of
describing biexciton formation.
At large $d$ there is a quantitative difference between the models
since the repulsive dipole-dipole interaction differs from the in-layer
Coulomb interaction.
We find that the bosonic dipole model gives condensate fractions which
are in good quantitative agreement with our results within the excitonic
phase for $r_s=7$--$8$~a.u.\ and $d>0.4$~a.u.

\begin{figure*}[htb!]
  \begin{center}
    \includegraphics[width=0.97\textwidth]{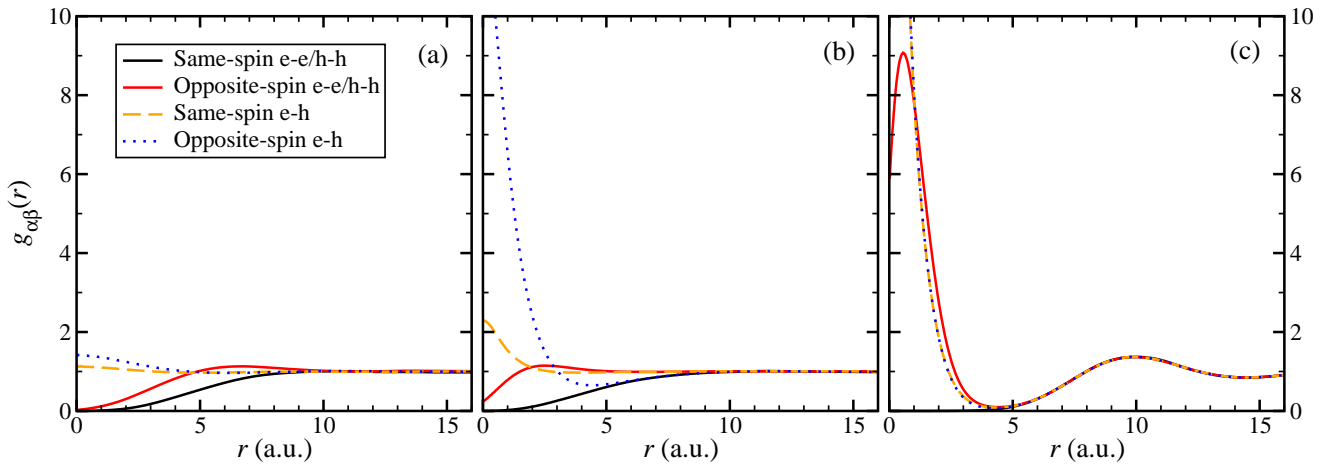}
  \end{center}
  \caption{
    (Color online)
    DMC PCFs as a function of the inter-particle distance $r$ at a
    density of $r_s=4$~a.u.\ and inter-layer separations of
    (a) $d=3$~a.u.\ in the fluid phase,
    (b) $d=0.4$~a.u.\ in the excitonic phase, and
    (c) $d=0$~a.u.\ in the biexcitonic phase.
    Note the strong enhancement of the opposite-spin e-h PCF at small
    $r$ in the excitonic phase, and the strong enhancement of both
    the opposite-spin and same spin e-h PCFs at small $r$ in the
    biexcitonic phase.}
  \label{fig:pcf_rs4}
\end{figure*}

The PCFs within the e-h bilayer at $r_s=4$ a.u.\ for $d = 0$, $0.4$,
and $3$~a.u.\ are shown in Fig.\ \ref{fig:pcf_rs4}.\cite{supplemental}
For $d=3$ a.u.\ the coupling between the layers is weak and the system
is in the fluid state.  The Fermi hole for same-spin e-e/h-h pairs is
wider than the correlation hole for opposite-spin e-e/h-h pairs.
The PCFs for same and opposite spin e-e/h-h correlations in
Fig.~\ref{fig:pcf_rs4}(a) are indistinguishable from those calculated
for the two-dimensional electron gas
($d\rightarrow\infty$).\cite{supplemental}
The enhancement of the same and opposite spin e-h PCFs at small $r$ is
very weak because of the large layer separation.

The PCF for antiparallel-spin e-h pairs at $d=0.4$ a.u., see
Fig.\ \ref{fig:pcf_rs4}(b), is strongly peaked at zero in-plane
separation, while the parallel-spin e-h PCF shows a shallow trough
with a PCF of $0.97$ at about $r=3$~a.u.\ and a small peak at
$r=0$~a.u.
The difference between these two PCFs is due to the fact that our
wave function explicitly binds antiparallel-spin electron-hole pairs.
The PCF for parallel-spin e-e/h-h pairs is strongly suppressed at
small separations, and is nearly identical to the parallel-spin
e-h/h-h PCF at $d=4$~a.u.
The PCF for antiparallel-spin e-e/h-h pairs shows a small peak
with a PCF of $1.15$ at $r\simeq 2.4$~a.u., significantly closer to
the origin than peak in the corresponding PCF at $d=4$~a.u.
There is almost no correlation hole in this PCF, reflecting the fact
that opposite-spin excitons are allowed to be close to each other.
The PCF is very close to unity for $r>10$~a.u.
The PCFs demonstrate the existence of an excitonic phase at
$d=0.4$~a.u.

The PCFs for $r_s=4$~a.u.\ and $d=0$ in the biexcitonic phase,
depicted in Fig.\ \ref{fig:pcf_rs4}(c), show very different features.
The PCFs show substantial long range oscillations which are not
present in the excitonic or one-component fluid phases.
The PCFs are strongly peaked at the origin for both parallel- and
antiparallel-spin e-h pairs, while the PCF for antiparallel-spin
e-e/h-h pairs shows a fairly strong peak and the PCF for parallel-spin
e-e/h-h pairs is close to zero for $r<0.4$ a.u.
Clearly the particles are aggregating into an object larger
than an exciton as the PCF for antiparallel-spin e-e/h-h pairs is
substantial at small $r$.
The fact that the parallel spin e-e/h-h PCF is essentially zero at
small $r$ tells us that the object in question contains, at most, one
particle of each type.
Direct integration of the PCFs confirms that the object contains one
particle of each of the four types, and that it is therefore a
biexciton.
The formation of objects larger than a biexciton is impeded by Pauli
exclusion.
Noting also the oscillations in the $d=0$ PCFs which decay with
distance, we can identify this phase as a biexcitonic fluid.
The diameter of the biexciton, measured as the median distance
between the anti-parallel-spin electrons, is about $1.46$~a.u.
At $r_s=6$ and $d=0$ we estimate the biexciton diameter to be
$1.48$~a.u.\cite{supplemental}

In summary, we use a wave function form of sufficient flexibility to
describe the fluid, excitonic and biexcitonic phases.
As the excitonic phase lies between the fluid and biexcitonic phases
we identify the phase transitions by the existence of a
non-zero excitonic condensate fraction, and the fluid and biexcitonic
phases can be distinguished by their characteristic PCFs.
The good agreement of our VMC and DMC expectation values suggests that
our results are of good quality, as does the agreement between the
results obtained using the SJ and SJB wave functions.
Excitons are unstable to biexciton formation at low densities and
$d<0.38$ a.u.\ in the bilayer system considered
here.\cite{Schindler_2008,Lee_2009}
For small values of $d$, we have found that biexcitons can survive
down to about $r_s=2.5$~a.u., which is a considerably higher density
than suggested previously.\cite{Zhu_1996_Physica_Scripta}

R.M.\ is grateful for financial support from KAKENHI grants 23104714
and 22104011, and from the Tokuyama Science Foundation.
P.L.R.\ and R.J.N.\ acknowledge financial support from the Engineering
and Physical Sciences Research Council (EPSRC) of the United Kingdom.
T.O.\ acknoledges support from the JSPS through its FIRST Program,
and from DYCE via KAKENHI grant 20104008.
The authors wish to thank Ms.\ Mitsumi Fujita for her assistance with
the calculations.


\end{document}